\documentclass[runningheads]{svmult}
\usepackage{makeidx}   
\usepackage{graphicx}  
\usepackage{subeqnar}  
\usepackage{multicol}  
\usepackage{physprbb}  
\makeindex             

\begin{document}
\title*{Progenitors of GRBs originated in the dense star clusters}
\toctitle{Progenitors of GRBs originated in the dense star clusters}
%
%
\titlerunning{Progenitors of GRBs}
%
\author{Yuri N. Efremov}
\authorrunning{Yu. N. Efremov}
%
%
\institute{Sternberg Astronomical Institute, 119899 Moscow, RF}

\maketitle              


\vspace{1cm}

The suggestion formulated in the title of this paper  arised first
to explain the origin of the giant (170 -- 300 pc in radii)
stellar arcs in the region of the supershell LMC4 in the LMC [4,5,7].
The formation of stellar arcs in the swept up gas shells
formed by the central sources of pressure needs
some $10^{52}$ ergs, yet neither multiple SNe in suggested central clusters
[6]
nor cloud  impacts  were able to form these arcs. I believe the first
suggestion failed to explain why are all the arcs so rare features,
all the arcs in the LMC  being close to each other, and the second
one -- why are the ages of arcs  different.

Yet another source of the energy imput to ISM to produce star-forming HI shells
was suggested to be the GRB events [8,15]
Along with the only known in the LMC  Soft Gamma
Repeater SGR 0526-66,  within the same region of $\sim$ 1 kpc in diameter
there are HI supershell and three or four  arc--shaped star complexes.
There should be in this region the common source for the progenitors of all
these objects! And there is indeed an unique star cluster in the same region:
the NGC 1978 cluster, 2 Gyr old and $10^6$ suns massive. There are no clusters
of similar mass and age  in the Milky Way galaxy, and only a handful of these
is in the LMC, NGC 1978 being the most massive and the most elongated one.

The binaries of compact objects, the progenitors of GRB,  might not be
results of two SN outbursts in the primeval binaries of massive stars.
Instead of this, the single stellar remnants could form  the hard binaries
in result of dynamical interactions in dense cluster cores. Many of them
might then have been dynamically ejected from the cluster, to merge after
escaping in GRB events, and to form the giant arcs not centered on NGC 1978
itself [4,7].

This  conclusion was supported by observation that X-ray binaries concentrate
near NGC 1978. The recent data for the LMC X-ray sources [19]
suggest  that in 10 $\times$ 10 degrees square there are nine X-ray binaries,
whereas  within 0.6 $\times$ 0.6  degrees square including NGC 1978,  there are
four of these nine stars.  It is tempting to  explain this with the origin
of these four X-ray binaries in NGC 1978, facing the high rate of occurence
of X-ray binaries in globular clusters and the high chance for them to be
dynamically ejected from the cluster (see refs. in [4,7]).

The recent studies of dynamical evolution of star clusters suggested
the high rate of the binary formation and star coalescences and ejections.
Portegies Zwart and MacMillan [18] have argued that the BH/BH binaries
are formed in (and many of these then ejected from) the massive clusters
in a few Gyr after a number of close  encounters. The ejected pairs  are
close enough to merge in a few Myr and the GRB events should be then rather
close to the parent cluster. The objects escaped from NGC 1978 must have
had quite reasonable spans of time  and velocities to merge in the centers
of the present day stellar arcs [5].

The possibility of the formation of compact binaries -- progenitors of
GRB in stellar encounters within the globular clusters was suggested also
by authors [14],
whereas authors [2]
considered
the generation of GRB due to the  stellar encounters in dense stellar
clusters of evolved galactic nuclei. Other arguments for the origin
of GRBs in star clusters were given by Kulkarni (this conference).

Bloom et al. [1] argued that  the GRB  afterglows observed mostly in
the regions of star formation (SFR) and therefore the GRB events are
connected with Hypernovae.  Even so, the progenitors of the latters
are suggested to  be the very massive fast rotating stars, and the most
plausible chanell of their formation is the coalescences of massive stars
inside  dense young clusters. There are indeed observational evidences
that hypernovae from  these stars did occur inside or near such clusters
in NGC 6946 [5,11]
and M82 galaxies  [16].

Anyway, the occurence of an afterglow near a SFR might just imply that this
SFR is the result of the previous  GRB event near the region, from the same
parent cluster, which might be rather old, like NGC 1978. The afterglow
distribution in Z suggests that most GRBs  have arised  8 -- 12 Gyrs ago,
being a few Gyrs younger than the systems of classical globulars in galaxies
like ours.  The delay of about 2 Gyrs is just the age of the NGC 1978 cluster.
It is even probable that this delay, which is close to the age difference
between the globular clusters  and the oldest objects of the Galactic disk,
implies that the formation of the massive stars in the disk was triggered
by the first GRB occured when the present day globulars were some 2 Gyr old
[5].

Contrary to authors [1] conclusions, the host-normalized offset
distribution of the GRBs seems to be similar nor to distribution of the star
formation regions neither to distribution of SNe [21].
In fact, it looks rather like the distribution of the classical (old) globular
clusters, with  the clear concentration to the center of the composite galaxy.
It may be also compatible with  the origin of GRB progenitors
in the very dense star clusters in galactic nuclei, suggested by
authors [2].
Note also, that  many GRBs had nor the observed afterglows neither the host
galaxies and some GRB afterglows were observed  near the cores of
elliptic-like galaxies.

There is the direct evidence for the occurence of GRB 980425, identified
with SN1998bw, in a cluster [13];
moreover, the authors [13] noted  near the SN the arc--shaped feature.
This GRB being by far the nearest one ($\sim$ 40 Mpc), only there the liner
resolution is high enough. Note also that both well observed SGRs in the
Galaxy, SGR 1906-20 and SGR 1900+14, are found to be in the dense young
clusters [17].

At any rate, assuming the giant stellar arcs were formed by GRB--connected
events, it is possible to get some conlusions on the geometry of jets.
The shapes and orientations of the LMC arcs  suggest they are the partiall
shells and cannot be results of isotropic bursts in non-uniform ISM [9].
The arcs might be
formed by the jets with the corresponding opening angle (60 -- 90 degrees),
what is compatible with Usov [22] model of GRB. Otherwise, there might exist
the multiprecessing and long--standing narrow jets (Fargion: [12] and
this conference). Their  working surface might fill up the partiall shells,
triggering star formation [10].
Note that the
long--standing precessing jet of SS433  has being formed the  HI bubble
during some 10000 years [3].
Otherwise, the jet instability
might result in the bow shock with the wide working surface, like it is
the case for the star--forming jets from AGN.

Note anyway, that the visible shapes of the triggered star complexes
which are intrisically the partiall shells, depend on their orientation
to the line of sight, whereas their intrinsic parameters  depend
on the properties of the target clouds to be swept up and compressed
into star-forming ones. Also, there was evidently no gas to form the
opposite-side stellar arcs in the LMC, considering their sizes.

\end{document}